\documentclass[%
 reprint,
superscriptaddress,
 amsmath,amssymb,
 aps,
]{revtex4-2}
\usepackage{graphicx}% Include figure files
\usepackage{dcolumn}% Align table columns on decimal point
\usepackage{bm}% bold math
\usepackage[free-standing-units=true]{siunitx} % for consistent handling of SI units
\usepackage{hyperref}% add hypertext capabilities
\usepackage{xcolor}

\usepackage{comment}

\begin{document}

\preprint{APS/123-QED}

\title{Hamiltonian reconstruction via ringdown dynamics}

\author{Vincent Dumont}
\email{vdumont@phys.ethz.ch}
\affiliation{Laboratory for Solid State Physics, ETH Zürich, CH-8093 Zürich, Switzerland}

\author{Markus Bestler}
\affiliation{Department of Physics, University of Konstanz, D-78457 Konstanz, Germany}

\author{Letizia Catalini}
\affiliation{Laboratory for Solid State Physics, ETH Zürich, CH-8093 Zürich, Switzerland}

\author{Gabriel Margiani}
\affiliation{Laboratory for Solid State Physics, ETH Zürich, CH-8093 Zürich, Switzerland}

\author{Oded Zilberberg}
\affiliation{Department of Physics, University of Konstanz, D-78457 Konstanz, Germany}

\author{Alexander Eichler}
\affiliation{Laboratory for Solid State Physics, ETH Zürich, CH-8093 Zürich, Switzerland}

\date{\today}% It is always \today, today,

\begin{abstract}
Many experimental techniques aim at determining the Hamiltonian of a given system. The Hamiltonian describes the system's evolution in the absence of dissipation, and is often central to control or interpret an experiment. Here, we theoretically propose and experimentally demonstrate a method for Hamiltonian reconstruction from measurements over a large area of phase space, overcoming the main limitation of previous techniques. A crucial ingredient for our method is the presence of dissipation, which enables sampling of the Hamiltonian through ringdown-type measurements. We apply the method to a driven-dissipative system -- a parametric oscillator -- observed in a rotating frame, and reconstruct the (quasi-)Hamiltonian of the system. Furthermore, we demonstrate that our method provides direct experimental access to the so-called symplectic norm of the stationary states of the system, which is tied to the particle- or hole-like nature of excitations of these states. In this way, we establish a method to unveil qualitative differences between the fluctuations around stabilized minima and maxima of the nonlinear out-of-equilibrium stationary states. Our method constitutes a versatile approach to characterize a wide class of driven-dissipative systems.
\end{abstract}

\maketitle

\section{Introduction}

The evolution of any physical system is governed by an interplay between conservative and nonconservative forces. The former are generated by a Hamiltonian `landscape', i.e., the sum of all energy terms within a closed system. Understanding and controlling an experiment usually require knowledge of the system's Hamiltonian. In a realistic setting, however, it can be difficult to compute the Hamiltonian from first principles, as it requires full insight into all microscopic constituents. Alternatively, extracting the effective Hamiltonian from a measurement provides direct access to the system dynamics even when its theoretical model is incomplete. Experimentalists therefore invest much effort in calibrating the parameters of their system in order to extract and fit their effective model~\cite{Cuairan_2022}. A prime example is the calibration of qubits, whose gate operations rely on precise Hamiltonian estimations~\cite{Bairey_2019,qi2019determining,Lin_2014,Siva_2023}.

A vast majority of systems are inherently `open', i.e., the dissipative coupling to an environment cannot be ignored. An open system experiences fluctuations that cause it to sample the available landscape over time. This allows estimating a system's Hamiltonian by measuring the probability of finding it in a certain state while it is subject to fluctuations~\cite{florin1998photonic,mccann1999thermally, Chan_2007, jeney2004mechanical,tanaka2013nanostructured,rondin2017direct,zijlstra2020transition, militaru2021escape}. This method was used in experimental studies of the escape dynamics~\cite{militaru2021escape} and Kramers turnover~\cite{rondin2017direct} of a particle trapped in an optical potential, investigations of the force fields acting on these particles~\cite{wu2009direct, perez2018high}, or the stability of coupled nonlinear systems~\cite{heugel2023proliferation,alvarez2023biased}. Probabilistic methods have the drawback that certain states are rarely explored. As such, the methods are often inadequate in situations where regions of interest (in state space) are separated by large energy differences. In addition, these methods are insensitive to temporal correlations in the system, which pertain to the notion of causality of excitations, as quantified by e.g. out-of-time-ordered correlators~\cite{swingle2018unscrambling,Blocher_2022}.

In the presence of dissipative coupling to an environment, it is also possible to probe a system's Hamiltonian from its deterministic relaxation. This becomes feasible when the measured variables of interest are large enough to neglect the impact of fluctuations. By initializing a system in a well-defined state using an external drive and then turning off the drive, the system decays into a stationary state due to dissipative coupling to a large and, most often, Markovian reservoir. By measuring such a `ringdown' into a stationary state (attractor), both the nonlinearity of a system close to a stable solution~\cite{Bachtold_RMP2022,Antoni2012Nonlinear,polunin2016characterization,Catalini_2021} and the curvature of engineered potential landscapes~\cite{morales2018coupling} have been measured in the absence of a drive.

In principle, such ringdown-type experiments can be extended to driven-dissipative systems. This is particularly interesting for driven nonlinear systems with multiple stable oscillation states. In a rotating frame, such oscillations can appear as stationary states within their respective basins of attraction~\cite{kozinsky2007basins,yan2022energy}. There, the system dynamics can be described by a rotating-frame quasi-Hamiltonian that, for instance, allows understanding out-of-equilibrium phase transitions~\cite{woo1971fluctuations,Dykman_1998,Soriente_2020,Soriente_2021,Ferri2021} in lattices of cold atoms~\cite{morales2018coupling}, optical oscillators~\cite{roy2021spectral}, or cavity magnonic systems~\cite{Zhang_2021}. Interestingly, such systems can stabilize out-of-equilibrium phases where seemingly anti-causal excitations can manifest~\cite{scarlatella2019spectralfunction,Soriente_2020}. However, to our knowledge, no full reconstruction of such a rotating-frame Hamiltonian has been reported to date.

In this paper, we demonstrate a precise and deterministic method of reconstructing the full rotating frame quasi-Hamiltonian of a driven-dissipative nonlinear system using systematic ringdown measurements. Notably, the presence of dissipation enables sampling of a large section of a Hamiltionian from a limited number of ringdown measurements. The method provides high resolution even far from stationary solutions, where stochastic approaches typically fail. Furthermore, we obtain a direct measurement of the `symplectic norm' of each solution~\cite{Soriente_2020,Soriente_2021}, providing a qualitative understanding of the different phases the system can enter, including a dissipation-stabilized maximum with hole-like quasiparticle excitations. Importantly, our method can be extended and applied to both undriven and driven-dissipative oscillating systems far from equilibrium, making it a valuable tool in many contemporary fields of physics.

We first present in Sec.~\ref{sec:device} our electro-mechanical resonator system. In Sec.~\ref{sec:HamiltonianReconstruction}, we introduce the general method to reconstruct the Hamiltonian from ringdown measurements. In Sec.~\ref{sec:HarmonicOscillator}, we reconstruct the effective Hamiltonian of our resonator in the absence of a drive -- i.e., demonstrating the method for a standard damped harmonic oscillator. Then, in Sec.~\ref{sec:ParametricOscillator}, we subject the resonator to a large parametric drive to allow multiple stable oscillation solutions and compare the reconstructed rotating-frame Hamiltonian to theoretical predictions. Finally, in Sec.~\ref{sec:Symplectic_Norm}, we discuss the extraction and implications of the symplectic norm as a tool for exploring and understanding out-of-equilibrium stationary solutions.

\section{Electromechanical Device}\label{sec:device}

\begin{figure}[!htb]
	\centering
	\includegraphics[width=0.95\columnwidth]{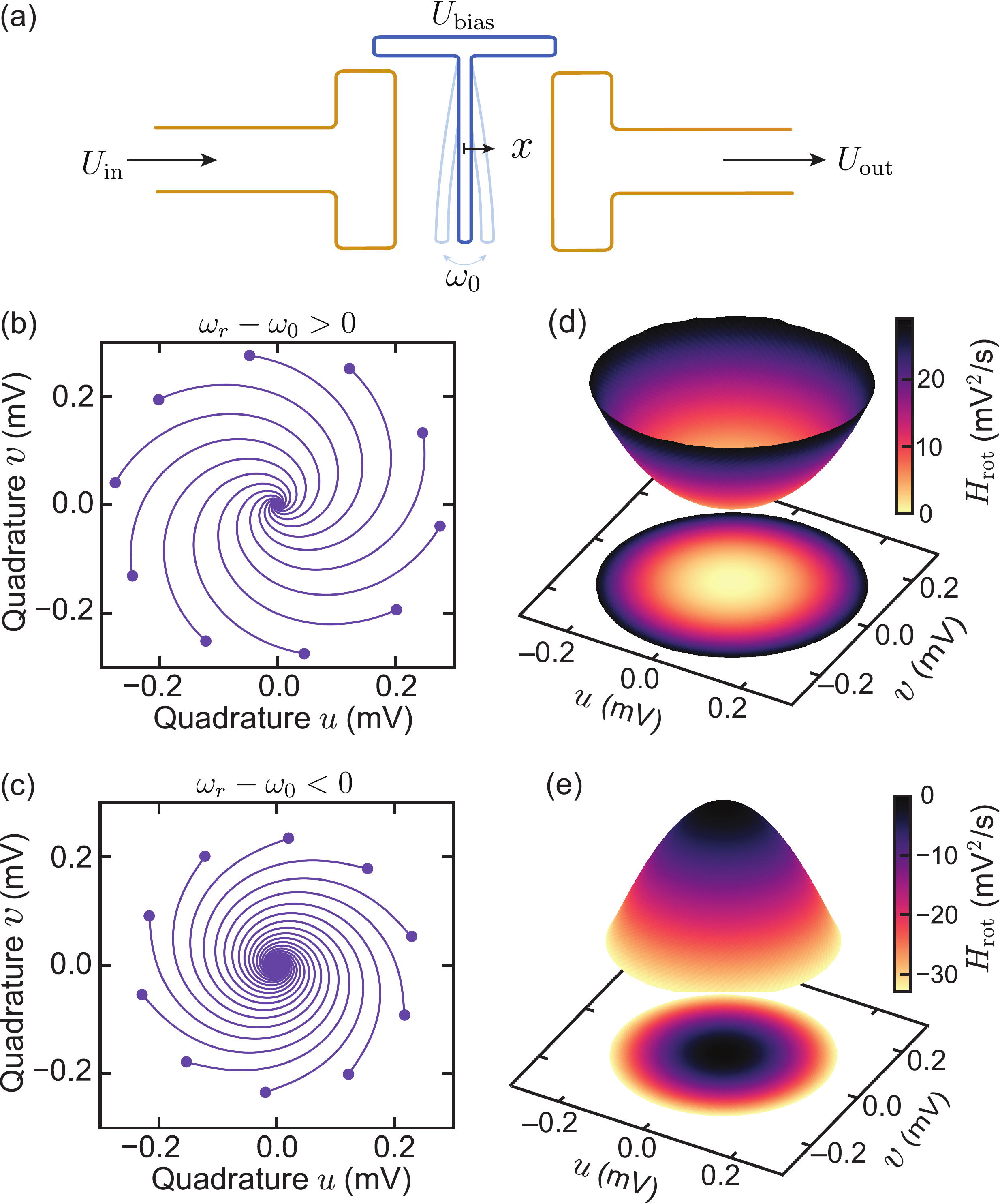}
	\caption{Micro-electromechanical resonator and $H_\mathrm{rot}$ reconstruction of a simple damped harmonic oscillator.  (a)~A cantilever, biased at voltage $U_\mathrm{bias}$, with oscillation displacement $x$ and natural frequency $\omega_0$, is capacitively coupled to two conductors. Typically, an input voltage $U_\mathrm{in}$ with frequency $\omega_\mathrm{r}$  ($2\omega_\mathrm{r}$) can drive the mechanical resonator resonantly (parametrically), when $\omega_\mathrm{r}\approx \omega_0$. The mechanical motion induces a voltage $U_\mathrm{out}$, which we read out. (b) Several ringdown measurements  for a few initialization points (dots) in rotating (i.e., at demodulated frequency $\omega_\mathrm{r}$) phase space $(u,v)$ for blue detuned demodulation, $(\omega_\mathrm{r} -\omega_0)/2\pi \approx   91$ Hz. (c)~Several ringdown measurements in phase space for red detuned demodulation, $(\omega_\mathrm{r} -\omega_0)/2\pi \approx -  209$ Hz. (d) Reconstructed Hamiltonian for positive detuning, i.e., from measurements of panel (b), using Eq.~\eqref{eq:Delta-H_rot}. (e) Reconstructed Hamiltonian for negative detuning, i.e. from the measurements of panel (c). }
	\label{Fig:1} 
\end{figure}

Our device is a micro-electromechanical resonator, as illustrated in Fig.~\ref{Fig:1}(a). It consists of a mechanical cantilever~\cite{agarwal2008study} biased with voltage $U_\mathrm{bias} \approx 32 $ V, used to tune the mechanical resonance frequency $\omega_0$~\cite{schmid2016fundamentals}, and which induces a Duffing non-linearity $\beta$ (due to the non-linearity of the electrostatic force between the mechanical element and the electrodes~\cite{agarwal2008study}). To generate a weak near-resonant forcing term $F$, we apply a voltage $U_\mathrm{in} = U_\mathrm{r} \cos(\omega_\mathrm{r} t + \theta)$ to one of the electrodes, where $U_\mathrm{r}$ is the amplitude, $\theta$ is a phase offset, and $\omega_\mathrm{r} \approx \omega_0$. A large off-resonant voltage $U_\mathrm{in} = U_\mathrm{p} \cos(2\omega_\mathrm{r} t + \psi)$ periodically modulates $\omega_0$ with a modulation depth $\lambda\propto U_\mathrm{p}$, making it possible to parametrically drive the resonator~\cite{Miller_2019_phase}. We read out the mechanical displacement $x$ via the output voltage $U_\mathrm{out} \propto x$ with a lock-in amplifier (Zurich Instruments MFLI), and forgo the proportionality coefficient for convenience (i.e., we define $U_\mathrm{out} \equiv x$)~\cite{Nosan_2019}. The device displacement is described by the equation of motion (EOM)
\begin{align}\label{eq:EOM}
    \frac{d^2x}{dt^2} + \omega_0^2[1-\lambda \cos(2\omega_\mathrm{r} t + \psi) ] x + \Gamma \frac{dx}{dt} + \beta x^3 =  F,
\end{align}
with time $t$. The device has a mechanical resonance frequency $\omega_0/2\pi \approx 1.12$ MHz, energy decay rate $\Gamma/2\pi \approx 112 $ Hz (quality factor $Q =\omega_0/\Gamma \approx 10^4$), and Duffing non-linearity $\beta \approx -9.9 \times 10^{16}$ V$^{-2}$s$^{-2}$. The forcing term (in units of Vs$^{-2}$) is $F = A U_\mathrm{in}$ with a conversion factor $A \approx 16 \times 10^6$ s$^{-2}$ and with the oscillator's mass absorbed in the definition of $F$. The parametric drive voltage amplitude $U_\mathrm{p}$ can be converted to a parametric modulation depth $\lambda = 2 U_\mathrm{p}/QU_{\mathrm{th}}$ by measuring the parametric drive threshold amplitude $U_{\mathrm{th}} \approx 1.98$ V. Above the parametric threshold, the system can be driven into an out-of-equilibrium stationary oscillation via a spontaneous time-translation symmetry breaking~\cite{Marthaler_2006, Ryvkine_2006, Chan_2007, Mahboob_2008, Wilson_2010, Leuch_2016, Gieseler_2012, Lin_2014, Grimm_2019, wang_2019, Miller_2019_phase, yamaji_2022}. The full device characterization is shown in Appendix~\ref{app:Exp_Charac}.

\section{Hamiltonian Reconstruction}
\label{sec:HamiltonianReconstruction}
We now introduce our method for reconstructing the Hamiltonian from measured coordinates along a trajectory.The evolution of a classical lossless system in the laboratory frame with coordinate $x$ and its conjugate $p$ is given by Hamilton's EOMs
\begin{align}\label{eq:H_SHO1D}
    \begin{aligned}
        \frac{dx}{dt} &= +\frac{\partial H(x,p)}{\partial p},  \\
        \frac{dp}{dt} &= -\frac{\partial H(x,p)}{\partial x},
    \end{aligned}
\end{align}
where $H(x,p)$ is the  Hamiltonian. Here, for any initial conditions and without drive, trajectories form closed loops along equipotential lines in coordinate space $(x,p)$ since there is no energy loss. This means that only a single loop of the Hamiltonian can be sampled when following any given trajectory. To sample all potential elevations, one has to initialize the system at infinitely many different initial conditions.

In many cases, we are interested in the slowly-varying in-phase and out-of-phase quadratures $u(t)$ and $v(t)$ observed in a frame rotating at $\omega_\mathrm{r} \approx \omega_0$, which are defined via $x(t) = u(t)\cos(\omega_\mathrm{r} t) - v(t) \sin(\omega_\mathrm{r} t)$. The demodulated quadrature we measure with our lock-in amplifier at frequency $\omega_\mathrm{r}$ are precisely $u$ and $v$. By applying the averaging method~\cite{eichler2023classical} on an equation of motion such as Eq.~\eqref{eq:EOM}, we obtain the so-called slow-flow equations~(see Appendix~\ref{app:Linearized-Model}). We open the system by adding  dissipative terms $\propto \Gamma$, i.e., we consider rotating Lagrangian dynamics,  such that the equations of motion  can be formulated in a similar structure as Eqs.~\eqref{eq:H_SHO1D}:
\begin{align}
\begin{aligned}\label{eq:du,dv}
        \frac{d u}{dt} &= + \frac{\partial H_\mathrm{rot}(u,v)}{\partial 
    v}  - \frac{\Gamma}{2} u, \\
    \frac{dv}{dt} &= - \frac{\partial H_\mathrm{rot}(u,v)}{\partial u}   - \frac{\Gamma}{2} v.
\end{aligned}
\end{align}
In this frame, the energy-conserving evolution of the system is governed by a rotating-frame quasi-Hamiltonian $H_\mathrm{rot}$.  For our resonator, subjected to both parametric and external drives, $H_\mathrm{rot}$ reads \footnote{Note that $H_\mathrm{rot}$ may not have dimensions of energy in this notation because $u$ and $v$ have the same units. The quasi-Hamiltonian can be expressed in units of energy by changing $H_\mathrm{rot}\rightarrow H_\mathrm{rot}/m\omega_\mathrm{r}$ in Eqs.~\eqref{eq:du,dv} and \eqref{eq:H_rot}.}
\begin{align}
\label{eq:H_rot}
    H_\mathrm{rot} = &  \frac{\omega_\mathrm{r}^2 - \omega_0^2}{4\omega_\mathrm{r}}(u^2 + v^2)\nonumber\\
    &-\frac{3\beta}{32\omega_\mathrm{r}}(2u^2v^2 + u^4 + v^4) \nonumber \\
    &+ \frac{\lambda \omega_0^2}{8\omega_\mathrm{r}}\left( 2uv\sin\psi  + (u^2 - v^2) \cos\psi\right)\nonumber\\
    &+ \frac{A U_\mathrm{r}}{2\omega_\mathrm{r}}\left( u\cos\theta + v\sin\theta \right).
\end{align}
 This Hamiltonian $H_\mathrm{rot}$ is a function in the rotating phase space spanned by $u$ and $v$~\cite{Bachtold_RMP2022}, and is the quantity we reconstruct in this paper. 

The addition of dissipation in Eq.~\eqref{eq:du,dv} is crucial. In the presence of dissipation, a trajectory is no longer confined to a single closed loop, but samples different energy elevations, ending up in one of the stable stationary states of the system, see Fig.~\ref{Fig:1}(b).  This allows us to probe $H_\mathrm{rot}$ by experimentally measuring $u(t)$ and $v(t)$ using rotating-frame ringdown measurements. By initializing the system in an initial state $(u_\mathrm{i}, v_\mathrm{i})$ and letting it evolve to a final state $(u_\mathrm{f}, v_\mathrm{f})$, we can extract the change in the Hamiltonian $\Delta H_\mathrm{rot}$ at any point $(u_j, v_j)$ along this ringdown's trajectory. We isolate $H_\mathrm{rot}$ in Eqs.~\eqref{eq:du,dv} and integrate over the slow coordinates from $(u_\mathrm{i}, v_\mathrm{i})$ to $(u_j, v_j)$ to obtain
\begin{align}\label{eq:Delta-H_rot}
     \Delta H_\mathrm{rot} = \int_{v_\mathrm{i}}^{v_j} \left[\frac{du}{dt} + \frac{\Gamma}{2} u \right] dv - \int_{u_\mathrm{i}}^{u_j} \left[\frac{dv}{dt} +\frac{ \Gamma }{2}v\right] du.  
\end{align}
In practice, the values $(u_j ,v_j)$ are measured in discrete steps, allowing to compute $\Delta H_\mathrm{rot}$ only at these points. Measuring multiple ringdowns with different initial $(u_\mathrm{i}, v_\mathrm{i})$, $\Delta H_\mathrm{rot}$ can be deterministically sampled and reconstructed over  a large area of phase space, with a resolution limited by the measurement uncertainty or fluctuations (e.g. thermal or quantum noise) in $u$ and $v$. Note that Eq.~\eqref{eq:Delta-H_rot} does not provide the relative change of $\Delta H_\mathrm{rot}$ between different ringdown measurements.

To compare different traces, we make use of the fact that $H_\mathrm{rot}(u_\mathrm{f}, v_\mathrm{f})$ should be single-valued at stationary points. This means that all traces sharing the same final coordinates $(u_\mathrm{f}, v_\mathrm{f})$ should have the same final value $H_\mathrm{rot}(u_\mathrm{f}, v_\mathrm{f})$. We thus find the relative Hamiltonian offset between different ringdown traces sharing the same $(u_\mathrm{f}, v_\mathrm{f})$ by comparing $H_\mathrm{rot}(u_\mathrm{f}, v_\mathrm{f})$. 
Finding the offset between traces is more complicated when they have different values of $(u_\mathrm{f}, v_\mathrm{f})$, i.e., they do not share a common end point. Here, the offset can be calculated by making the Hamiltonian continuous, i.e. by finding the offset that minimizes the difference between nearby starting points $(u_\mathrm{i}, v_\mathrm{i})$ that decay into different final $(u_\mathrm{f}, v_\mathrm{f})$, see Appendix~\ref{App:Hamiltonian_Reconstruction}. 

Note that a Hamiltonian reconstruction analogous to Eq.~\eqref{eq:Delta-H_rot} can also be performed in the nonrotating frame, i.e., for a dissipative Eq.~\eqref{eq:H_SHO1D}. We concentrate here on the case of a rotating-frame Hamiltonian to be in line with the theory. In the following, we will test the Hamiltonian reconstruction before discussing what information we can extract from it.

\section{Harmonic Oscillator Case}\label{sec:HarmonicOscillator}

As a first demonstration, we reconstruct the rotating-frame Hamiltonian of a damped harmonic oscillator. In Figs.~\ref{Fig:1}(b) and (c), we show multiple measured trajectories for the case $\lambda = 0$, and with amplitudes that are small enough to neglect the effect of the Duffing non-linearity, i.e. only considering the first line in Eq.~\eqref{eq:H_rot}. For each of those ringdown trajectories, we first displace the resonator in phase space using a near-resonant drive with fixed $U_\mathrm{r}$ and $\omega_\mathrm{r}$, and with an individually selected $\theta$. In a frame rotating at $\omega_\mathrm{r}$, the corresponding initial resonator coordinates $(u_\mathrm{i}, v_\mathrm{i})$, shown as dots, are stationary under the resonant drive. Then, we switch off the drive  ($U_\mathrm{r} = F=0$), and track the ringdown trajectory while the system decays to the state $(u_\mathrm{f}=0, v_\mathrm{f}=0)$. During this decay, the angle of the state in phase space evolves in time since the rotating frame's frequency is detuned from the resonance frequency~\cite{Antoni2012Nonlinear}. For $\omega_\mathrm{r} - \omega_0 > 0$ in Fig.~\ref{Fig:1}(b), the trajectories spiral towards $(0, 0)$ with a clockwise orientation, while for $\omega_\mathrm{r} - \omega_0 < 0$ in Fig.~\ref{Fig:1}(c), they spiral in counterclockwise direction.

We reconstruct the rotating-frame Hamiltonian $H_\mathrm{rot}$ from the ringdown trajectories using Eq.~\eqref{eq:Delta-H_rot}. For the reconstructions in Fig.~\ref{Fig:1}(d) and (e), the resolution was improved by using more ringdown traces than shown in Figs.~\ref{Fig:1}(b) and (c). The resulting $H_\mathrm{rot}$ is a paraboloid whose sign of curvature depends on  $\omega_\mathrm{r} - \omega_0$. This is predicted in Eq.~\eqref{eq:H_rot} and reflects the fact that $H_\mathrm{rot}$ describes energy relative to a rotating frame. Indeed, for $\omega_\mathrm{r} - \omega_0 = 0$, the rotating potential would be entirely flat. Note that the origin $(0,0)$, which appears as a maximum of $H_\mathrm{rot}$ in Fig.~\ref{Fig:1}(e), remains the only stable solution of the system irrespective of the detuning. As a (local) maximum can appear as a stationary state, this example demonstrates that the rotating-frame Hamiltonian cannot be interpreted as a simple energy function as is the case in the non-rotating frame~\cite{eichler2023classical}.
After successfully reconstructing the rotating frame potential of a harmonic oscillator, we now apply our method to a more complex system.

\section{Parametric Oscillator Case}\label{sec:ParametricOscillator}

The reconstruction of $H_\mathrm{rot}$ can also be applied to driven nonlinear systems with multiple stable solutions. Here, we demonstrate this principle on the example of the electromechanical resonator described in Sec.~\ref{sec:device} when subjected to a parametric drive, cf. Eq.~\eqref{eq:H_rot} and Fig.~\ref{Fig:1}(a).

\begin{figure}[t]
	\centering 
	\includegraphics[width=0.9\columnwidth]{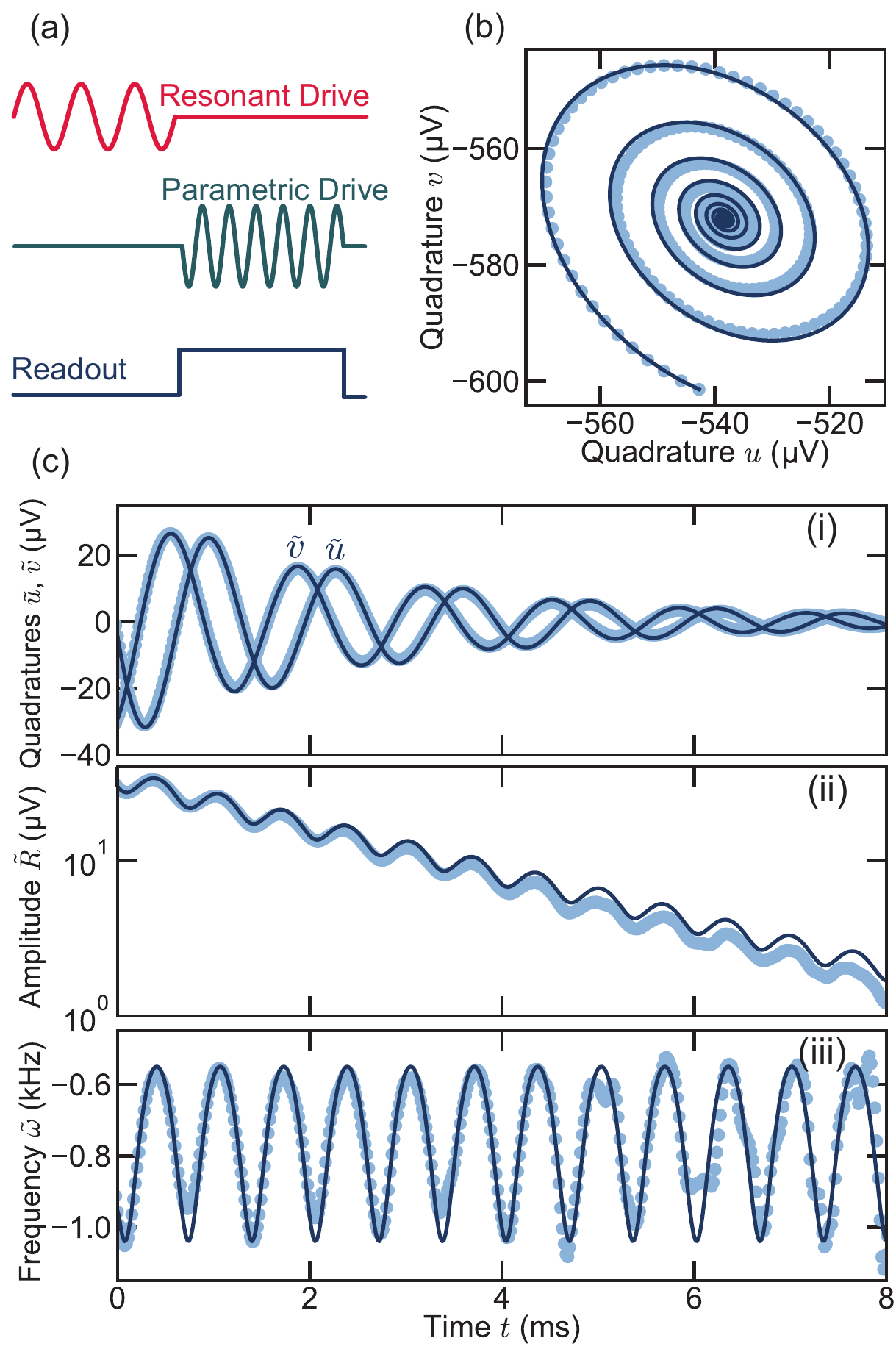}
	\caption{Ringdown of a parametric oscillator. (a)~For a ringdown measurement, the resonator is first displaced with a near-resonant drive $U_\mathrm{in}(\omega_\mathrm{r})$ with $\omega_\mathrm{r}\approx \omega_0$. This resonant drive is turned off and a parametric drive $U_\mathrm{in}(2\omega_\mathrm{r})$ is immediately turned on, while the slowly-varying (i.e., demodulated at $\omega_\mathrm{r}$) quadratures $u$ and $v$ are read out. (b)~Ringdown in phase space for fixed parametric drive $U_\mathrm{p} = 10$ V ($\lambda \approx 10^{-3} $) and frequency $(\omega_\mathrm{r} - \omega_0)/2\pi \approx  -253$ Hz. The dark blue curve is the predicted ringdown with the linearized model in Appendix~\ref{app:Linearized-Model} [Eq.~\eqref{eq:LinearizedModelSolutionLinearizedEquationOfMotion}]. (c)~In panels (i) the in-phase $\tilde{u}$ and out-of-phase  $\tilde{v}$ displaced quadratures,  (ii) amplitude $ \tilde{R} = \sqrt{\tilde{u}^2 + \tilde{v}^2}$, and (iii) instantaneous rotating-frame frequency $\tilde{\omega}= \partial_t\tilde{\phi}/2\pi $ with phase $\tilde{\phi}=\arctan(\tilde{v}/\tilde{u})$ are plotted as a function of time for the same ringdown as in (b). These measurements are referenced to the attractor in which they ringdown into, i.e., $\tilde{u} \equiv u - u_\mathrm{f}$ with $u_\mathrm{f} \approx \SI{-538}{\micro\volt}$ the final value of $u$ (the position of the attractor), and similarly for $\tilde{v}$ with  $v_\mathrm{f} \approx \SI{-572}{\micro\volt}$. }
	\label{Fig:2} 
\end{figure}

We first analyze a single rotating-frame ringdown into an out-of-equilibrium stationary state in phase space. In Fig.~\ref{Fig:2}, we study an example trajectory of our system in the presence of parametric driving beyond the instability threshold ($\lambda > \lambda_\mathrm{th}$)~\cite{eichler2023classical}, where two time-translation symmetry-broken high-amplitude solutions are stabilized due to the interplay between the drive and the nonlinearity, see Appendix \ref{app:Exp_Charac}. As shown schematically in Fig.~\ref{Fig:2}(a), the system is initialized in $(u_\mathrm{i}, v_\mathrm{i})$ by a near-resonant force at $\omega_\mathrm{r}$, followed by a parametric drive tone whose amplitude $\lambda \propto U_\mathrm{p}$ and frequency $2\omega_\mathrm{r}$ define the stationary solutions of the system in the readout frame rotating at frequency $\omega_\mathrm{r}$~\cite{Lifshitz_Cross,DykmanBook,eichler2023classical}. In Fig.~\ref{Fig:2}(b), we show a typical ringdown trajectory of our resonator in phase space, noting that $(u_\mathrm{f}\neq 0, v_\mathrm{f}\neq 0)$.

To facilitate the Hamiltonian reconstruction, we define displaced quadratures $\tilde{u} \equiv u - u_\mathrm{f}$ and $\tilde{v} \equiv v - v_\mathrm{f}$, with the final coordinates of the ringdown  $(u_\mathrm{f},v_\mathrm{f})$ (i.e, the attractor position). In this shifted reference frame, the relaxation into the stable solution resembles a ringdown process in an equilibrium system, with $\tilde{u}$ and $\tilde{v}$ performing damped oscillations towards a final state $(\tilde{u}_\mathrm{f}=0,\tilde{v}_\mathrm{f}=0)$, see Fig.~\ref{Fig:2}(c)(i). The damping rate associated with these trajectories is quantified by the same $\Gamma$ as we introduced in Eq.~\eqref{eq:EOM}, while the rotation around $(u_\mathrm{f}, v_\mathrm{f})$ depends on $H_\mathrm{rot}$. The dynamics of these shifted quadratures are well captured by linearizing the equations of motion [Eqs.~\eqref{eq:du,dv}] around the attractors, see derivation in Appendix~\ref{app:Linearized-Model}.

\begin{figure*}[!htb]
	\centering
	\includegraphics[width=0.85\textwidth]{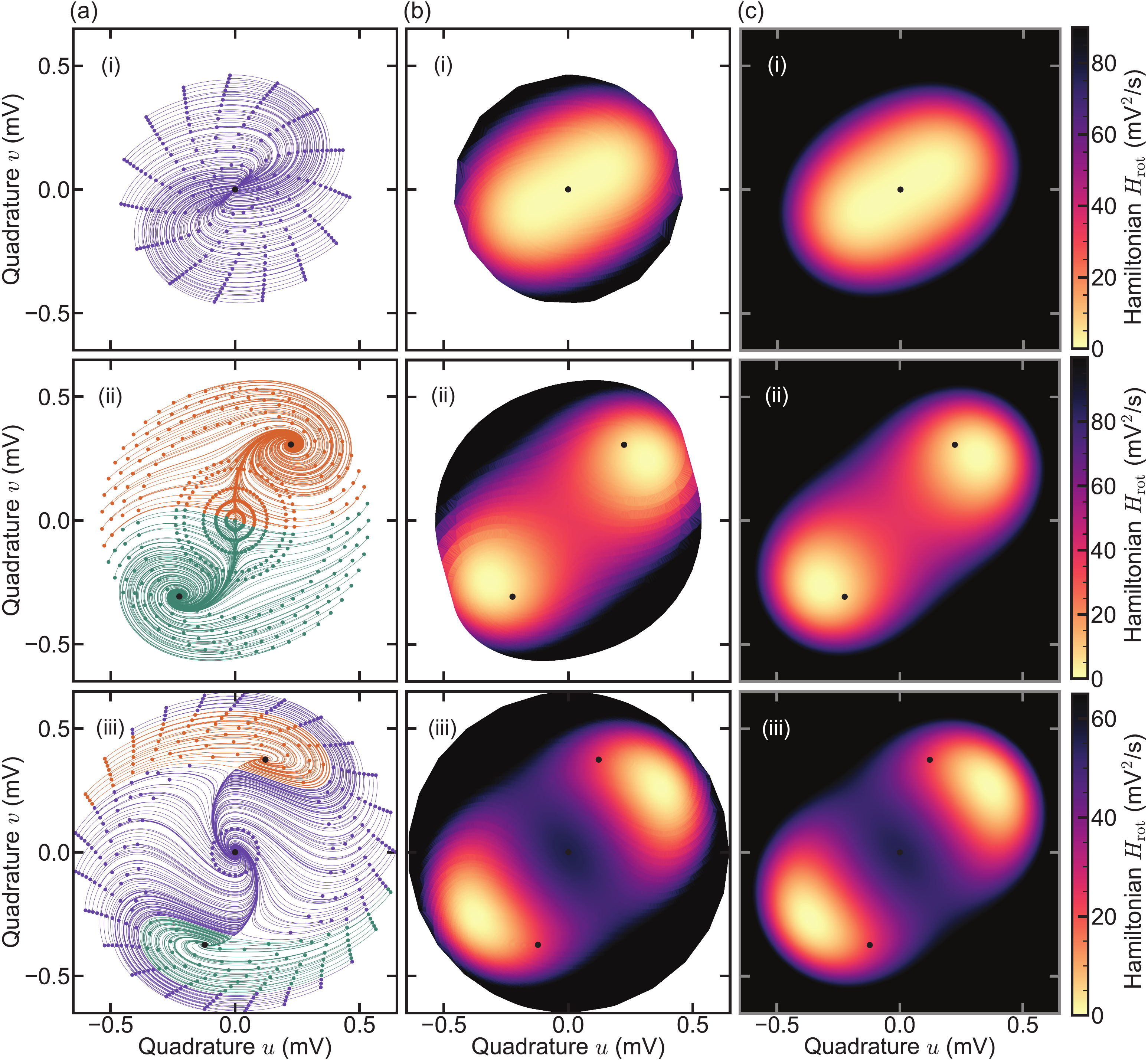}
	\caption{Ringdowns and Hamiltonian reconstruction in parametric phase space. (a)~Multiple ringdowns in phase space under different parametric drive strenght and frequency. Each ringdown's starting point is indicated by a dot, and color-coded for the attractor in which it rings down: purple the zero-amplitude, orange the top-right, and green the bottom-left attractor. The attractors extracted from the ringdown measurements are indicated by black dots. The parametric drive is (i)~below threshold, (ii)~above threshold, and (iii)~above threshold and far detuned, cf. Appendix \ref{app:Exp_Charac} for the locations of the measurements relative to the device's phase diagram. (b)~Rotating-frame Hamiltonian reconstruction from the ringdown measurements in  (a), cf. Eq.~\eqref{eq:Delta-H_rot}. (c)~Theoretical plots  of the rotating-frame Hamiltonian [Eq.~\eqref{eq:H_rot}] using potential angle (i)~$\psi \approx -2.1$ rad, (ii)~$\psi \approx -1.76$ rad, and (iii)~$\psi \approx -1.86$ rad as the only fit parameter, while fixing the independently measured mechanical parameters, see Appendix~\ref{app:Exp_Charac}. Note that the attractors are not necessarily at the minima of the quasi-Hamiltonian due to dissipation.}
	\label{Fig:3} 
\end{figure*}

In contrast to undriven relaxation processes in the lab frame $(x, p)$, the amplitude $\tilde{R} = \sqrt{\tilde{u}^2 + \tilde{v}^2}$ does not decay monotonically. Instead, we observe oscillations imposed on top of the exponential decay in Fig.~\ref{Fig:2}(c)(ii). These oscillations stem from the fact that the shape of $H_\mathrm{rot}$ around $(u_\mathrm{f}, v_\mathrm{f})$ is not rotationally symmetric, such that the system samples different $\partial_u H_\mathrm{rot}$ and  $\partial_v H_\mathrm{rot}$ as it moves around the Hamiltonian landscape. We emphasize that despite the fact that we observe  oscillations growing in amplitude  at certain times, this does not violate any law of conservation, as we are dealing with a driven system. The non-rotationally symmetric Hamiltonian also manifests in Fig.~\ref{Fig:2}(c)(iii), where an instantaneous rotating-frame frequency $\tilde{\omega}= \partial_t \tilde{\phi}/2\pi$ is defined by the phase $\tilde{\phi}=\arctan(\tilde{v}/\tilde{u})$ relative to the attractor. 

There are clear oscillations in the frequency $\tilde{\omega}$ due to the  resonator sampling different $\partial_u H_\mathrm{rot}$ and  $\partial_v H_\mathrm{rot}$ as it moves in phase space. This behavior is also well captured by the averaged and linearized dynamics  
near the attractors (cf. Eqs.~\eqref{eq:du,dv} and Appendix~\ref{app:Linearized-Model}). Note that for smaller parametric drives or for dynamics further away from the attractors, one needs to consider a model that goes  beyond the linearization we employ (i.e. expand the equations to higher-order terms) to fully capture the dynamics.

In a next step, we measure multiple trajectories with various initialization conditions, but fixed parametric drive strength and frequency, allowing us to reconstruct the rotating frame Hamiltonian $H_\mathrm{rot}$ of our driven nonlinear system. In Fig.~\ref{Fig:3}(a), we show such sets of trajectories for three different parametric drives: (i)~below threshold with a single squeezed state, (ii)~above threshold with two stable phase states, and (iii)~at large driving and large detuning with a combination of phase states and a zero-amplitude state~(see Appendix~\ref{app:Exp_Charac} for a phase diagram of the device). Each ringdown is color-coded by the  state it eventually approaches, allowing us to identify the corresponding attractor pools. From ringdown measurements, we can thus directly identify the number of stable solutions in the given phase-space area, as well as the separatrices of the system -- the regions where the color of nearby ringdown measurements changes. In addition, we obtain a visualization of the stream flow in phase space, as governed by Eq.~\eqref{eq:du,dv}. We stress that the main assumption of this reconstruction is that the damping is linear, cf. Eqs.~\eqref{eq:du,dv}, though these equations could be adapted to include other types of damping. The reconstruction does not assume linearized dynamics.

The reconstructed Hamiltonians $H_\mathrm{rot}$ are shown in Fig.~\ref{Fig:3}(b) alongside theoretical Hamiltonians in Fig.~\ref{Fig:3}(c), as calculated from Eq.~\eqref{eq:H_rot} with independently measured parameters $\omega_0$, $\beta$, and $\lambda$, see Appendix~\ref{app:Exp_Charac}. We find excellent qualitative agreement between measurement and theory. Crucially, the quality and resolution of the Hamiltonian reconstruction is consistently high over the entire sampled phase space, which would be hard to achieve with statistical methods~\cite{florin1998photonic,mccann1999thermally, Chan_2007, jeney2004mechanical,tanaka2013nanostructured,rondin2017direct,zijlstra2020transition, militaru2021escape}. In the reconstructed $H_\mathrm{rot}$, we can clearly see the appearance of one, two, and three stable states in the three cases (i)-(iii), respectively, indicating different phases of the driven system~\cite{woo1971fluctuations,Dykman_1998,Heo_2010}. Our reconstruction method confirms the theoretical prediction that the stationary solutions can qualitatively differ in the rotating frame; while the phase states at finite amplitude in the cases (ii) and (iii) are marked by a minimum in $H_\mathrm{rot}$, the stable state appearing at $(u=0, v=0)$ in (iii) corresponds to a maximum, signalling a fundamentally different type of solution, i.e., a dissipation-stabilized state~\cite{Soriente_2020, Soriente_2021}. In the following, we analyze this difference using the symplectic norm of the individual solutions, and we show that this quantity yields valuable insights into the behavior of driven-dissipative systems.

\section{Symplectic Norm}\label{sec:Symplectic_Norm}

In sections~\ref{sec:HarmonicOscillator} and~\ref{sec:ParametricOscillator}, we successfully reconstructed rotating-frame Hamiltonians $H_\mathrm{rot}$ using ringdown measurements. One of the most prominent distinctions of a rotating-frame Hamiltonian, compared with a Hamiltonian in a laboratory frame such as in Eq.~\eqref{eq:H_SHO1D}, is that both maxima and minima of $H_\mathrm{rot}$ can constitute stable oscillation states of the system, unlike what we expect from the minimal action principle in equilibrium systems. This counter-intuitive feature can be clearly observed in row (iii) of Fig.~\ref{Fig:3}, where two minima at finite amplitudes are separated by a stable local maximum at $(u=v=0)$. Importantly, the minima in Fig.~\ref{Fig:3} are stable due to the nonlinearity $\beta$, while the maximum is stabilized by dissipation~\cite{Soriente_2021}. This fundamental difference, however, is difficult to quantify in standard measurements, such as frequency sweeps and stability diagrams~\cite{Mahboob_2008,karabalin2010efficient}.

A method tailored to classify and distinguish minima and maxima in $H_\mathrm{rot}$ is the symplectic norm $ds^2$~\cite{Soriente_2020, Soriente_2021}. This quantity indicates if the excitations (i.e., out-of-equilibrium  phonons) of a system around an attractor are more hole-like or particle-like: when an excitation with a negative (positive) symplectic norm is created on top of a stationary system, it reduces (increases) the energy of the system relative to the rotating frame, that is, relative to an excitation at the driving frequency $\omega_\mathrm{r}$. This difference manifests in the Hamiltonian: stationary solutions with a negative (positive) symplectic norm appear as maxima (minima) in $H_\mathrm{rot}$ and are formally associated with a hole-like (particle-like) excitation, see Appendix~\ref{app:Linearized-Model}. We can therefore use the reconstructed Hamiltonians in Fig.~\ref{Fig:3} to directly determine the symplectic norm of the stable oscillations states, and to classify the corresponding excitations.

A careful study of Fig.~\ref{Fig:3} reveals additional information. Namely, the trajectories leading to Hamiltonian extrema assume clockwise or counter-clockwise rotations. For our system, we show in Appendix~\ref{app:Linearized-Model} that the symplectic norm is directly linked to the sense of rotation of the trajectories close to an attractor. This allows us to extract the symplectic norm $ds^2$ of the different attractors directly from the measured ringdown trajectories, even without a full reconstruction. To capture this observation mathematically, we define a correlator characterizing the direction of rotation of a ringdown,
\begin{align}
\label{eq:GreensFunction}
    G^c(t'-t)=\Theta(t'-t)\langle\tilde v(t) \tilde u(t')-\tilde u(t)\tilde v(t')\rangle_t,
\end{align}
with the Heavyside step-function $\Theta(t-t')$, and $\langle ...\rangle_t$ indicating an average over all times $t$. This quantity $G^c(t-t')$ is a classical analogue to the quantum Green's function used in  Refs.~\cite{soriente2018dissipation,Soriente_2021,scarlatella2019spectralfunction}. Following these works, we calculate the corresponding spectral response $\mathcal{A}$ of a stable oscillation state
\begin{align}
    \mathcal A(\omega) &= -2\mathrm{Im}[G^c(\omega)] \label{eq:A_exp}\\
    &= \frac{ |\zeta|^2 ds^2}{2} \left[\frac{1}{(\omega-\omega_\mathrm{lin})^2+\frac{\Gamma^2}{4}}-\frac{1}{(\omega+\omega_\mathrm{lin})^2+\frac{\Gamma^2}{4}}\right]\nonumber  ,
\end{align}
where $G^c(\omega)$ is the Fourier transform of $G^c(t-t')$, $\zeta$ is a constant related to the ringdown starting conditions, and $\omega_\mathrm{lin}$ is the oscillation frequency of the quadratures. The linear response function in the second line is only valid near the attractor of interest.

Looking at $\mathcal{A}$, we see that the spectral response has peaks at $\pm\omega_\mathrm{lin}$, with width $\Gamma$, and an overall sign which is determined by the symplectic norm $ds^2$.
Therefore, if the resonator rings down with a counter-clockwise (clockwise) rotation in phase space, the final steady state of the system has a negative (positive) symplectic norm and a negative (positive) peak in $\mathcal{A}$.

\begin{figure}[t]
	\centering
	\includegraphics[width=0.75\columnwidth]{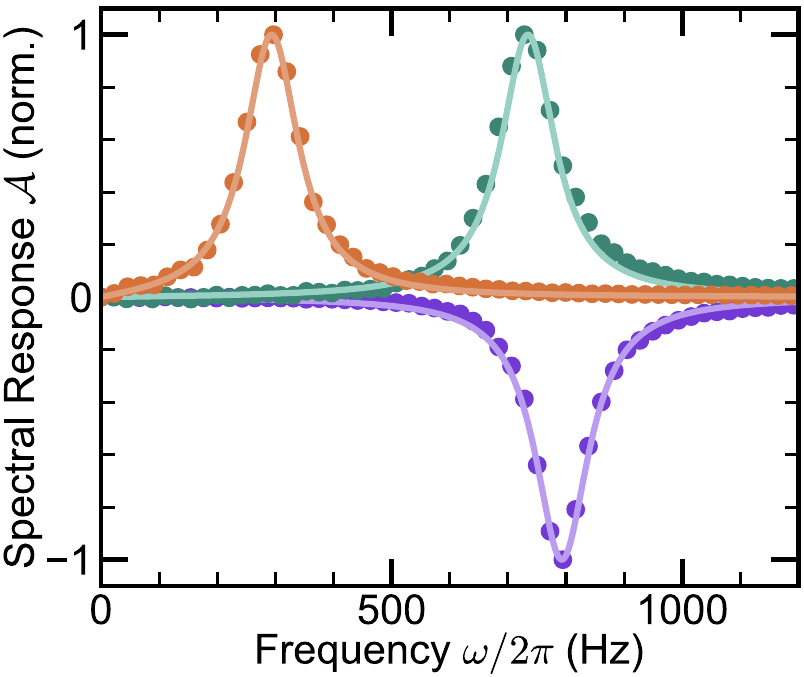}
	\caption{Spectral response of a parametric oscillator. With $\tilde{u}$ and $\tilde{v}$ obtained during a ringdown measurement, we can extract the spectral response $\mathcal{A}$ using Eqs.~\eqref{eq:GreensFunction} and \eqref{eq:A_exp} (dots) [cf. Figs.~\eqref{Fig:2} and \eqref{Fig:3}]. These spectral responses are in good agreement with the fitted theoretical curves of Eq.~\eqref{eq:A_exp}, with $\omega_0$ left as a free parameter to account for small frequency drifts. All measurements are done with the same parametric drive strenght $U_\mathrm{p} = 10$ V ($\lambda \approx 10^{-4}$), and different demodulation frequencies $\omega_\mathrm{r}$. The detuning is $(\omega_\mathrm{r} - \omega_0)/2\pi = 170 $ Hz (orange) and  $(\omega_\mathrm{r} - \omega_0)/2\pi = - 222 $ Hz (green) for ringdowns into high amplitude states, and  $(\omega_\mathrm{r} - \omega_0)/2\pi =  -865 $ Hz (purple) for the zero amplitude state.}
	\label{Fig:4} 
\end{figure}

With this knowledge, we can extract the sign of $ds^2$ from the spectral response of ringdown measurements, as shown in Fig.~\ref{Fig:4}. We record $\tilde{u}$ and $\tilde{v}$ during one trajectory, calculate the correlator $G^c(t-t')$ [Eq.~\eqref{eq:GreensFunction}], take its Fourier transform, and compute the spectral response $\mathcal{A}$ according to Eq.~\eqref{eq:A_exp}. In Fig.~\ref{Fig:4}, we present the measured result for each of the three stable states of a parametric oscillator at different detunings. We compare these measured spectral responses with those expected from theory, calculated using independently measured parameters  and the linearized slow-flow equations [Eq.~\eqref{eq:A_exp}], and obtain very good agreement. As expected, for a stable solution corresponding to a maximum in $H_\mathrm{rot}$ (purple curve), the spectral response is a dip, while the two other solutions, which are minima in $H_\mathrm{rot}$ (orange and green curves), correspond to peaks in the spectral response. These measurements show the link between the orientation of rotation, the symplectic norm, the spectral function, and the rotating frame Hamiltonian $H_\mathrm{rot}$. 

We emphasize that the notion of maxima and minima of the rotating quasienergy potential depend on the chosen rotating frame frequency, and that stabilized maxima are a manifestation of out-of-equilibrium stationary states. As such, the fact that the excitations on top of the maxima are hole-like implies that their dynamics are slower than the clock in the rotating frame as manifested by the chirality of their ringdown. In other words, we can think of the response in this case as non-causal relative to the clock. This is analogous to how antiparticles, that exhibit a negative mass dispersion, behave in relativistic quantum mechanics~\cite{dirac1932relativistic}.

\section{Outlook}
We report a precise method to reconstruct the Hamiltonian of a system via ringdown measurements. The method allows for a full characterization of the energy potential, including multiple stable solutions, saddle points, and attractor pools. The method is particularly suited for studies in the growing field  of driven-dissipative nonlinear systems, where a Hamiltonian characterization from first principles is often very difficult. Furthermore, it bestows the ability to characterize the symplectic norm of different stable oscillation solutions in the rotating frame with a connection to relativistic quantum mechanics and causality. We expect that this approach will allow the classification of a broad variety of systems, including nanomechanics, superconducting circuits, light-matter systems, and nonlinear optics.

\section*{Acknowledgements}
We acknowledge Christian Marti and Sebastián Guerrero for help building the measurement setup. We thank  Nicholas E. Bousse and T. W. Kenny for providing the MEMS device, and Matteo Fadel and Tobias Donner for useful discussions. V. D. acknowledges support from the ETH Zurich Postdoctoral Fellowship Grant No.~23-1 FEL-023. O. Z. acknowledges funding from the Deutsche Forschungsgemeinschaft (DFG)
via project number 449653034 and through SFB143. A. E. and O. Z. acknowledge financial support from the Swiss National Science Foundation (SNSF) through the Sinergia Grant No.~CRSII5\_206008/1.

%\pagebreak
\appendix

\section{Experimental Characterization} \label{app:Exp_Charac}

\subsection{Parametric Sweeps}

In this section, we describe how we measure the parametric response of our resonator in order to extract its resonance frequency $\omega_0$, energy decay rate $\Gamma$, and Duffing nonlinearity $\beta$. 

\begin{figure}[t]
	\centering
	\includegraphics[width=0.85\columnwidth]{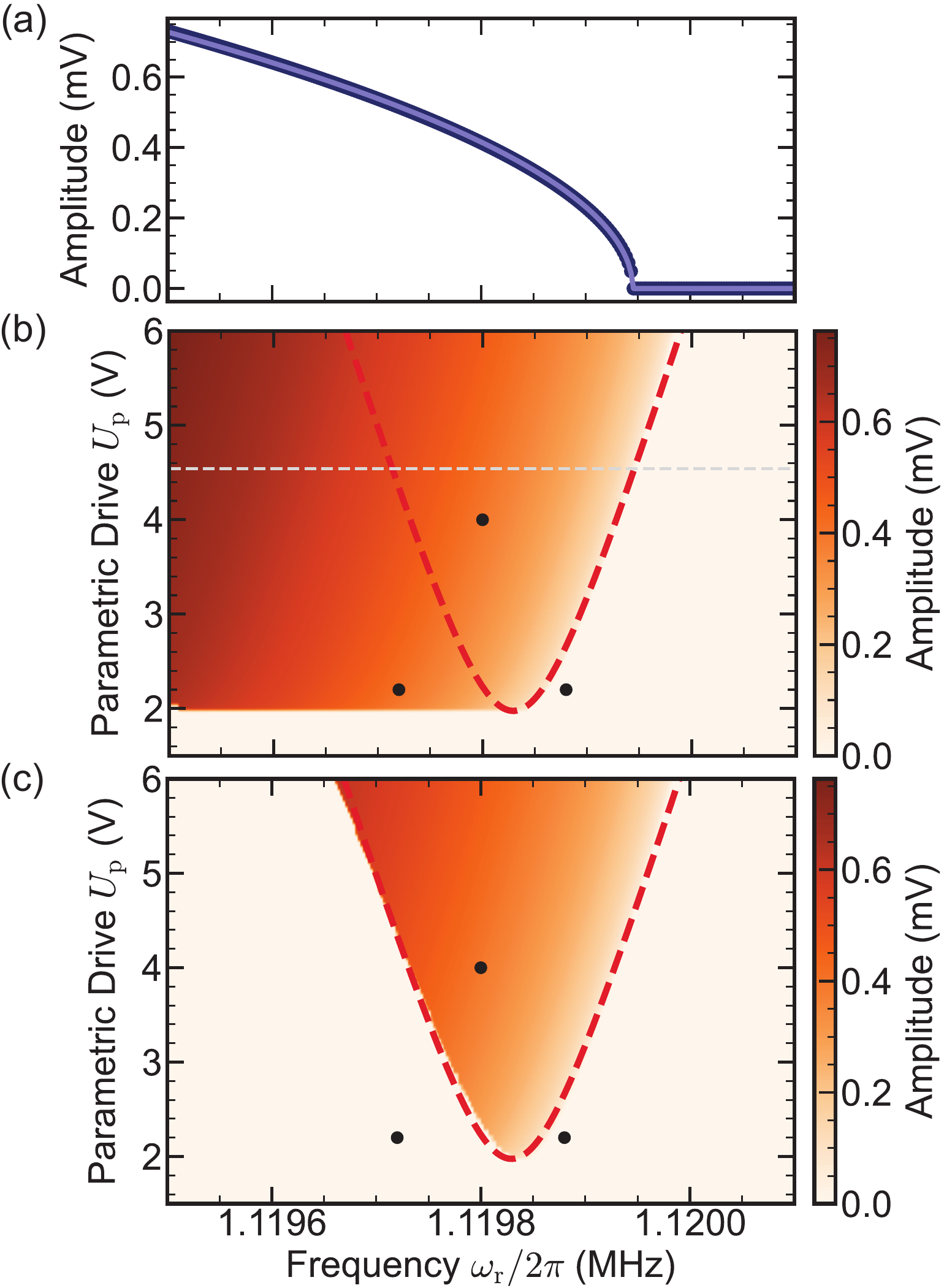}
	\caption{Parametric response.  (a)~Response of our resonator to a parametric drive $U_\mathrm{in} = U_\mathrm{p} \cos(2\omega_\mathrm{r} t)$ with voltage amplitude $U_\mathrm{in} = 4.54$ V. The frequency is swept from high to low values. (b)~Response to a parametric drive as in (a) for different values of $U_\mathrm{p}$. The dashed grey line indicates the measurement of panel (a), the red dashed line corresponds to the fitted `Arnold tongue' outline [see panel (c)], while the dots indicate the parameters used in Fig.~\ref{Fig:3}. (c)~Response to a parametric drive as in (a) for different values of $U_\mathrm{p}$ when sweeping from low to high frequencies.
 }
	\label{Fig_app:1} 
\end{figure}

We send a parametric tone  $U_\mathrm{in} = U_\mathrm{p} \cos(2\omega_\mathrm{r} t)$ to our device and read out the induced displacement at frequency $\omega_\mathrm{r}$.
In Fig.~\ref{Fig_app:1}(a), we show the system's response while sweeping the frequency of the parametric drive from high to low frequencies for fixed parametric drive voltage $U_\mathrm{p} = 4.54$~V. By repeating this measurement for different parametric drive voltages, we obtain the `phase diagram' of our parametric oscillator, shown in Fig.~\ref{Fig_app:1}(b). Now, performing the same measurement but instead sweeping the frequency from low to high frequencies (i.e., ``against'' the Duffing nonlinearity), we obtain a different diagram commonly referred to as `Arnold Tongue', shown in Fig.~\ref{Fig_app:1}(c). 

Comparing with theoretical predictions~\cite{eichler2023classical}, Figs.~\ref{Fig_app:1}(b) and (c) allow us to read out the different phases and number of solutions of our resonator depending on the parametric drive strength and frequency. Indeed, the outline of the Arnold tongue, shown as a red dashed line, indicates the (frequency-dependent) parametric threshold, above which the resonator has exactly two stable states (the parametric phase states), both of which have finite-amplitude but opposite phases. The orange region outside the Arnold tongue in Fig.~\ref{Fig_app:1}(b) features a zero-amplitude state in addition to the two parametric phase states. Which state is selected depends on the initial condition of the resonator. Finally, in the white region in Fig.~\ref{Fig_app:1}(b) only the zero-amplitude state is stable. 

To find the equation predicting the parametric response of our device, we use the `slow-flow' equations [cf.~Eqs.~\eqref{eq:du,dv} and~\eqref{eq:LinearizedModelSlowFlowEquations}], and find the steady state amplitude response $R= \sqrt{u^2 + v^2}$  by setting $\dot{u} = \dot{v}=0$, yielding~\cite{eichler2023classical}
\begin{align}\label{eq:amplitude_parametric}
     \left[ -\frac{\lambda^2}{4} + \left( \frac{\Gamma\omega}{\omega_0^2} \right)^2 + \left(1 - \frac{\omega^2}{\omega_0^2} + \frac{3}{4} \frac{\beta R^2}{\omega_0^2}\right)^2\right]R^2 = 0\,.
\end{align}
Equation~\eqref{eq:amplitude_parametric} has the trivial solution $R=0$. For the case $R\neq 0$, we can divide Eq.~\eqref{eq:amplitude_parametric} by $R^2$ to obtain the non-trivial solutions
\begin{align}\label{app:eq-ParamResp}
    R(\omega) = \left( \frac{4 \omega_0^2}{3\beta}  \left[  \left( \frac{\omega^2}{\omega_0^2}-1\right) \pm \left(\frac{\lambda^2}{4} - \frac{\Gamma^2 \omega^2}{\omega_0^4}\right)^{1/2} \right]\right)^{1/2}\,,
\end{align}
when the radicands of both square roots are positive. Thus, the parametric amplitude response allows (amongst other things) to extract the Duffing nonlinearity.

The outline of the Arnold tongue corresponds to the limit $R\to 0$ in Eq.~\eqref{app:eq-ParamResp}, leading to
\begin{align}
    \left( \omega^2 - \omega_0^2 \right)^2 - \frac{\lambda^2\omega_0^4}{4} + \Gamma^2 \omega^2  = 0 \,.
\end{align}
Solving for the parametric drive yields
\begin{align}\label{app:eq-Arnold}
    \lambda =   2 \sqrt{\frac{\Gamma^2\omega^2}{\omega_0^4} + \left( 1 - \frac{\omega^2}{\omega_0^2}\right)^2},
\end{align}
where we kept the (physical) positive solution.

To extract our device parameters, we start by fitting Eq.~\eqref{app:eq-Arnold} to the outline of the Arnold tongue, see red dashed line in Fig.~\ref{Fig_app:1}(c). For the fit, we replace $\lambda = U_\mathrm{p}/C$, where $C$ is a constant converting the unitless parametric drive strength $\lambda$ to the applied parametric voltage $U_\mathrm{p}$. Doing so, we obtain the resonance frequency $\omega_0/2\pi = 1.1198294 (3)$~MHz and the energy decay rate $\Gamma/2\pi = 112 (2)$~Hz of our resonator, as well as the conversion constant $C = 9.88 (5)\times 10^3$~V. 

In a second step, we use the measured parametric response in Fig.~\ref{Fig_app:1}(a) and fit it to Eq.~\eqref{app:eq-ParamResp} in order to extract the Duffing nonlinearity $\beta = -9.894(4)\times 10^{16}$ (V$\cdot$s)$^{-2}$, while fixing $\omega_0$, $\Gamma$, and $C$ to the previously fitted values. 

\subsection{Resonant Sweep}

We now extract the factor $A$, converting from an input voltage $U_\mathrm{in} = U_\mathrm{r}\cos(\omega_\mathrm{r} t)$ to an applied force $F$ (in units Vs$^{-2}$), from the resonant response of our resonator.

To calibrate this, we consider the EOM for our resonator under near-resonant (but small, i.e., neglecting the Duffing term) drive [cf. Eq.~\eqref{eq:EOM}],
\begin{align}\label{App:eq-x-EOM}
    \ddot{x}(t) + \omega_0^2 x(t) + \Gamma \dot{x}(t) =  A U_\mathrm{in}(t),
\end{align}
where we replaced the forcing term by $F= A U_\mathrm{in}$. Taking the Fourier transform of Eq.~\eqref{App:eq-x-EOM} yields
\begin{align}\label{App:eq-x-omega}
  x(\omega) =  A U_\mathrm{in}(\omega) \left[ \omega_0^2 -\omega^2  - i\omega \Gamma \right] ^{-1} ,
\end{align}
which provides a direct link between the voltage $U_\mathrm{in}$ applied to our device and the  voltage $U_\mathrm{out}\equiv x$ we read out.

\begin{figure}[t]
	\centering
	\includegraphics[width=0.75\columnwidth]{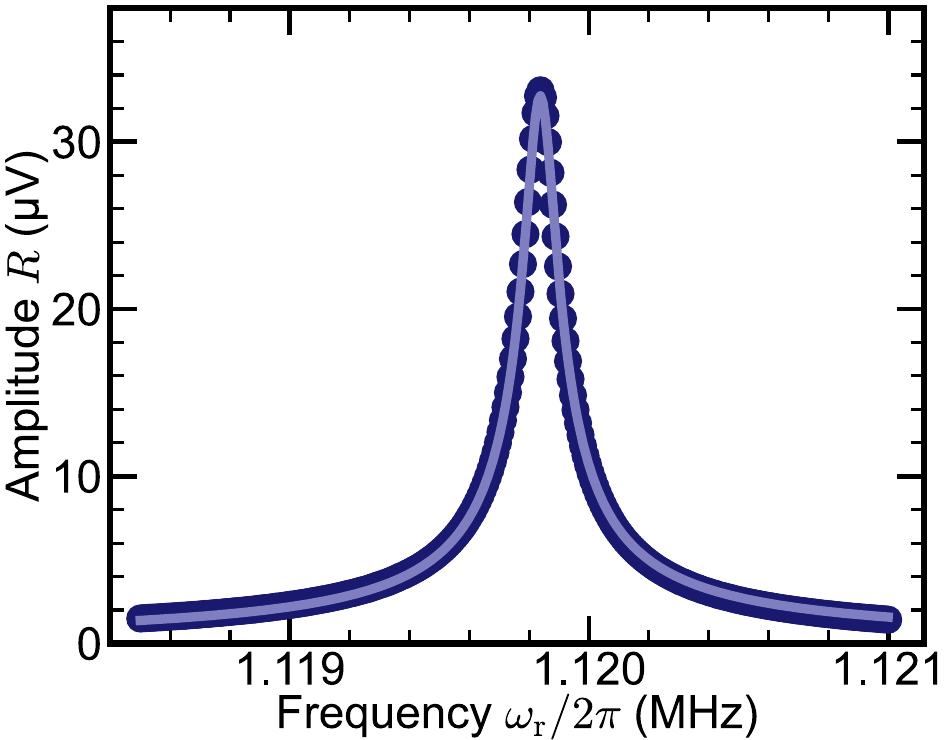}
	\caption{Linear resonant response. An input voltage $U_\mathrm{in} = U_\mathrm{r}\cos(\omega_\mathrm{r} t)$ with $U_\mathrm{r} = 10$~mV is swept across the resonance frequency and the amplitude response $R = \sqrt{u^2 +v^2}$ at frequency $\omega_\mathrm{r}$ is measured (dots). The data is fitted (line) according to Eq.~\eqref{App:eq-x-omega} with fixed $\Gamma/2\pi = 112$~Hz, while $\omega_0$ is left as a free parameter to account for small frequency drifts. We extract a value $A = 16.22(1) \times 10^6$~s$^{-2}$.}
	\label{Fig_app:2} 
\end{figure}

We can thus extract $A$ by sweeping the frequency $\omega_\mathrm{r}$ of a driving tone across the mechanical resonance of our resonator and measuring its response. In Fig.~\ref{Fig_app:2}, we fit the measured response with Eq.~\eqref{App:eq-x-omega} and obtain $A =16.22 (1) \times 10^6$~s$^{-2}$.

\section{Linearized Model}\label{app:Linearized-Model}
In this Appendix, we describe the method we use to obtain the EOMs for the rotating frame coordinates $u$ and $v$, and the link between the symplectic norm and the ringdown trajectories. We first obtain the steady state solutions under parametric drive, then we linearize the EOMs in their vicinity. We solve the linearized EOMs to get analytical expressions for the resonators rotation frequency and direction around the attractors. The latter reveals a direct link between the ringdown rotation and the symplectic norm of the corresponding final steady state. Finally, we show how to extract the symplectic norm from a ringdown measurement using the spectral response function.

\subsection{Averaging Method}
As described in the main text, starting from Eq.~\eqref{eq:EOM}, we move to a rotating frame with out-of-phase quadratures $u(t)$ and $v(t)$, defined via $x(t) = u(t)\cos(\omega_\mathrm{r}t)-v(t)\sin(\omega_\mathrm{r}t)$. Assuming that $u(t)$ and $v(t)$ vary slowly in time compared to the mechanical oscillations with frequency $\omega_\mathrm{r}$, we average the EOMs over the time $2\pi/\omega_\mathrm{r}$ to obtain approximated EOMs for $u(t)$ and $v(t)$~\cite{eichler2023classical},
\begin{align}
\label{eq:LinearizedModelSlowFlowEquations}
        \dot{u}&=-\frac{\Gamma}{2}u-\delta v-\frac{3\beta}{8\omega_\mathrm{r}}\left(u^2 v+ v^3\right)-\frac{\lambda\omega_0^2}{4\omega_\mathrm{r}}v\, ,\nonumber \\
        \dot{v}&=-\frac{\Gamma}{2}v+\delta u+\frac{3\beta}{8\omega_\mathrm{r}}\left(u^3+ u v^2\right)-\frac{\lambda\omega_0^2}{4\omega_\mathrm{r}}u\, ,
\end{align}
where we introduce the detuning $\delta \equiv (\omega_0^2-\omega_\mathrm{r}^2)/2\omega_\mathrm{r}$. For simplicity, we also set $\psi=0$ when moving from Eq.~\eqref{eq:EOM} to Eq.~\eqref{eq:LinearizedModelSlowFlowEquations}. This phase $\psi$ rotates the potential in phase space around the origin with an angle of $\psi/2$, so the solutions for $u$ or $v$ calculated for $\psi=0$ in this section can be rotated by an angle of $\psi/2$ to match the experimental results.

\subsection{Linearization of the EOMs}
To predict the motion of the resonator near an attractor, we linearize the slow-flow equations [Eqs.~\eqref{eq:LinearizedModelSlowFlowEquations}] near the attractors \cite{eichler2023classical, ricardo2023linearize}. To do so, we first determine the coordinates of the attractors (i.e., the stationary states of the system) in the rotating frame by setting $\dot{u}=\dot{v}=0$ in Eq.~\eqref{eq:LinearizedModelSlowFlowEquations}. Solving the resulting coupled polynomial equations leads to five possible solutions \cite{Kosata2022HarmonicBalance, Borovik2024HarmonicBalanceMaths}:
\begin{widetext}
\begin{align}
    (u_\mathrm{f},v_\mathrm{f})_1&=(0,0)\, ,\nonumber \\
    (u_\mathrm{f},v_\mathrm{f})_{2,3}&=\left(\pm u_{+},\pm \frac{u_+ ( 3\beta\lambda\omega_0^2 u_{+}^2 - 2\lambda^2\omega_0^4+4\Gamma^2\omega_\mathrm{r}^2+4\lambda\omega_0^4-4\lambda\omega_0^2\omega_\mathrm{r}^2}{2\Gamma\omega_\mathrm{r}\left(2\omega_\mathrm{r}^2-2\omega_0^2+\lambda\omega_0^2\right)}\right)\, ,\nonumber\\
    (u_\mathrm{f},v_\mathrm{f})_{4,5}&=\left(\pm u_{-},\pm \frac{u_- ( 3\beta\lambda\omega_0^2 u_{-}^2 - 2\lambda^2\omega_0^4+4\Gamma^2\omega_\mathrm{r}^2+4\lambda\omega_0^4-4\lambda\omega_0^2\omega_\mathrm{r}^2}{2\Gamma\omega_\mathrm{r}\left(2\omega_\mathrm{r}^2-2\omega_0^2+\lambda\omega_0^2\right)}\right)\, ,
    %(u_f,v_f)&=\left(\pm_1 u_{\pm_2},\mp_1 \frac{3\alpha\lambda\omega_0^2 u_{\mp_2}u_{\pm_2}^2}{2\Gamma\omega_\mathrm{r}\left(2\omega_\mathrm{r}^2-2\omega_0^2+\lambda\omega_0^2\right)}\right)
\end{align}
where
\begin{align}
    u_\pm&=\left [\frac{\beta\lambda\omega_0^2\left( 2\lambda\omega_0^2\omega_\mathrm{r}^2+\lambda^2\omega_0^4-2\lambda\omega_0^4-4\Gamma^2\omega_\mathrm{r}^2\right ) \pm\sqrt{\beta^2\lambda^2\omega_0^4\left(2\omega_\mathrm{r}^2-2\omega_0^2+\lambda\omega_0^2\right)^2\left(\lambda^2\omega_0^4-4\Gamma^2\omega_\mathrm{r}^2\right)}}{3\beta^2\lambda^2\omega_0^4}\right ]^{1/2} \, .
\end{align}
\end{widetext}
For any given set of parameters, a maximum of three of these solutions are stable, i.e., they act as attractors, cf. Fig.~\ref{Fig:3}. To see if a solution is stable, we now introduce the displaced quadratures $(\tilde u,\tilde v)_k\equiv(u-u_\mathrm{f},v-v_\mathrm{f})_k$, relative to a given solution position $(u_\mathrm{f},v_\mathrm{f})_k$, %For the non-trivial attractors,
and linearize Eq.~\eqref{eq:LinearizedModelSlowFlowEquations} around one attractor by neglecting higher-order terms in $\tilde{u}$ and $\tilde{v}$. This procedure yields
\begin{align}
    \label{eq:LinearizedModelEOM}
    \begin{pmatrix}
    \dot{\tilde{u}}\\
    \dot{\tilde{v}}    
    \end{pmatrix}
    &= 
    \underbrace{
    \left.
    \begin{pmatrix}
    \frac{\partial \dot{u}}{\partial u} & \frac{\partial \dot{u}}{\partial v}\\[0.15cm]
    \frac{\partial \dot{v}}{\partial u} & \frac{\partial \dot{v}}{\partial v}
    \end{pmatrix}
    \right|_{(u,v)=(u_\mathrm{f},v_\mathrm{f})_k}
    }_{\mathbf{J}_\mathrm{f}}
    \begin{pmatrix}
    \tilde{u}\\
    \tilde{v}    
    \end{pmatrix} \, .
\end{align}
Here, $\mathbf{J}_\mathrm{f}$ is the Jacobian of the slow-flow equations evaluated around the solution position $(u_\mathrm{f}, v_\mathrm{f})_k$, which takes the form
\begin{align}
    \mathbf{J}_\mathrm{f}
    &=
    \begin{pmatrix}
        -\frac{\Gamma}{2}-\frac{3\beta u_\mathrm{f} v_\mathrm{f}}{4\omega_\mathrm{f}} & -\delta-\frac{3\beta\left(u_\mathrm{f}^2+3v_\mathrm{f}^2\right)}{8\omega_\mathrm{r}}-\frac{\lambda\omega_0^2}{4\omega_\mathrm{r}}\\
        \delta +\frac{3\beta\left(3u_\mathrm{f}^2+v_\mathrm{f}^2\right)}{8\omega_\mathrm{r}}-\frac{\lambda\omega_0^2}{4\omega_\mathrm{r}} & -\frac{\Gamma}{2}+\frac{3\beta u_\mathrm{f} v_\mathrm{f}}{4\omega_\mathrm{r}}
    \end{pmatrix}
.
\end{align}
The Jacobian $\mathbf{J}_\mathrm{f}$ describes the linearized forces acting in the rotating frame near a given solution. Solving the linear first-order differential equation given by Eq.~\eqref{eq:LinearizedModelEOM} leads to
\begin{align}
\label{eq:LinearizedModelSolutionLinearizedEquationOfMotion}
    \begin{pmatrix}
        \tilde u\\
        \tilde v
    \end{pmatrix}
    =
    \zeta \mathbf{w}_+ e^{\mu_+ t}
    + \zeta^* \mathbf{w}_- e^{\mu_- t}\, .
\end{align}
Here,  
 \begin{align}
     \zeta =& \frac{1}{2}(v_\mathrm{i}-v_\mathrm{f})-i\bigg(\frac{3\beta u_\mathrm{f} v_\mathrm{f}}{8\omega_\mathrm{r}\omega_\mathrm{lin}}(v_\mathrm{i}-v_\mathrm{f})
     \\
     &+\frac{(9 \beta u_\mathrm{f}^2+3 \beta v_\mathrm{f}^2+4\omega_0^2-4\omega_\mathrm{r}^2-2\lambda\omega_0^2)}{16\omega_\mathrm{r}\omega_\mathrm{lin}} (u_\mathrm{i}-u_\mathrm{f})\bigg) \nonumber 
 \end{align}
is a constant which depends on the initial position $(u_\mathrm{i},v_\mathrm{i})$ of the resonator in the rotating frame. Furthermore, we use in Eq.~\eqref{eq:LinearizedModelSolutionLinearizedEquationOfMotion} the Jacobian's eigenvectors
\begin{align} \label{eq:solvedLinearized}
    \mathbf{w}_\pm = \begin{pmatrix}
        \frac{-6\beta u_\mathrm{f} v_\mathrm{f}\pm 8i\omega_\mathrm{r}\omega_\mathrm{lin}}{9 \beta u_\mathrm{f}^2+3 \beta v_\mathrm{f}^2+4\omega_0^2-4\omega_\mathrm{r}^2-2\lambda\omega_0^2}
        \\
        1
    \end{pmatrix},
\end{align}
and corresponding eigenvalues
\begin{align}\label{eq:eigenvalues}
    \mu_\pm = - \Gamma/2 \pm i\omega_\mathrm{lin},
\end{align}
 with the complex frequency
 \begin{align}
    \omega_\mathrm{lin}=\frac{1}{8\omega_\mathrm{r}}  \Bigl[&27\beta^2 u_\mathrm{f}^4 +6\beta u_\mathrm{f}^2\left(9\beta v_\mathrm{f}^2+8\omega_0^2-8\omega_\mathrm{r}^2+2\lambda\omega_0^2\right)
    \nonumber
    \\
    &+\left(3\beta v_\mathrm{f}^2+4\omega_0^2-4\omega_\mathrm{r}^2-2\lambda\omega_0^2\right)\nonumber\\ 
    &\times \left(9 \beta v_\mathrm{f}^2+4\omega_0^2-4\omega_\mathrm{r}^2+2\lambda\omega_0^2\right)  \Bigl]^{1/2} \, .
\end{align}
If the real part of one eigenvalue in Eq.~\eqref{eq:eigenvalues} is positive, Eq.~\eqref{eq:LinearizedModelSolutionLinearizedEquationOfMotion} diverges for $t\rightarrow\infty$ and $(u_\mathrm{f},v_\mathrm{f})_k$ is an unstable state. If both real parts are negative, $(\tilde u,\tilde v)$ converge to $(0,0)$ and $(u_\mathrm{f},v_\mathrm{f})_k$ is an attractor \cite{eichler2023classical, ricardo2023linearize}. The theory results depicted in Fig.~\ref{Fig:2}(b), (c)(i), and (c)(ii) are directly calculated using Eq.~\eqref{eq:LinearizedModelSolutionLinearizedEquationOfMotion}. 

The theory prediction for the frequency of the rotation around an attractor in Fig.~\ref{Fig:2}(c)(iii) is analytically calculated using
\begin{align}
\label{eq:LinearizedModelFrequency}
    \frac{1}{2\pi}\frac{d \tilde{\phi}}{d t}
    &=\frac{\partial}{\partial t}\arctan\left(\frac{\tilde{v}}{\tilde{u}}\right) 
    \nonumber\\
    &=\frac{1}{2\pi}\frac{\tilde{u}(d\tilde{v}/d t)-\tilde{v}(d\tilde{u}/d t)}{\tilde{u}^2+\tilde{v}^2}
    \nonumber\\
    &=-\frac{|\zeta|^2 \omega_\mathrm{lin}}{\pi} \frac{ds^2}{\tilde{u}^2+\tilde{v}^2}e^{-\Gamma t} \, ,
\end{align}
with the `symplectic norm' defined as
\begin{align}
\label{eq:SymplecticNormDefinition}
    ds^2 \equiv i(w_{+,1}w_{-,2}-w_{-,1}w_{+,2})\, ,
\end{align}
and where $w_{\pm,l}$ is the $l$-th entry of $\mathbf{w}_\pm$. The sign of $ds^2$ therefore decides the rotational sense of the ringdown. We now show that this quantity is identical to the symplectic norm derived for quantum driven-dissipative systems, which allows to classify  stable states in driven-dissipative systems~\cite{Soriente_2020, Soriente_2021, Fan2023symplectic}.

\subsection{Symplectic Norm}
To see that the symplectic norm [cf.~Eq.~\eqref{eq:SymplecticNormDefinition}] is the same as the one defined in quantum driven-dissipative systems, we transform our definition of the symplectic norm to the form commonly used in quantum optics~\cite{Soriente_2020, Soriente_2021, Fan2023symplectic}. First, we express the rotating frame coordinates $\tilde{u}(t)$ and $\tilde{v}(t)$ in terms of the complex coordinate $\alpha(t)$ and its complex conjugate $\alpha^*(t)$ via the transformation matrix $\mathbf{S}$:
\begin{align} \label{eq:S_transform}
    \begin{pmatrix}
        \tilde u
        \\
        \tilde v
    \end{pmatrix}
    &=
    \underbrace{\sqrt{\frac{\hbar}{2m\omega_0}}
    \begin{pmatrix}
        1 & 1
        \\
        i & -i
    \end{pmatrix}}_{\equiv \mathbf{S}^{-1}}
    \begin{pmatrix}
        \alpha
        \\
        \alpha^*
    \end{pmatrix}    \,.
\end{align}
This basis change can be interpreted as moving to the mean-field limit of bosonic creation and annihilation operators of a quantum harmonic oscillator representing the quadratures $\tilde{u}(t)$ and $\tilde{v}(t)$ (i.e., $\alpha=\langle a\rangle$ with a bosonic annihilation operator $a$ and expectation value $\langle...\rangle$). 
 
 Note that the $\sqrt{\hbar/2m\omega_0}$ prefactor is the same for both quadratures due to our definition of $\tilde{u}$ and $\tilde{v}$ sharing the same units. This prefactor leads to
 \begin{align}
    \label{eq:SNotUnitary}
     \mathbf{S}^{-1}=\frac{\hbar}{m\omega_0}\mathbf{S}^\dag,
 \end{align}
implying that $\mathbf{S}$ is not unitary and therefore not a norm-preserving transformation. This will later be accounted for by scaling the vectors in the new basis, and has no consequences for the sign of the symplectic norm, which is the relevant quantity for us.

We use the transformation introduced in Eqs.~\eqref{eq:LinearizedModelEOM} and~\eqref{eq:S_transform} to find the equations of motion for $\alpha$ and $\alpha^*$:
\begin{align}
\label{eq:DynamicMatrix}
    i\,
    \mathbf{S}
    \begin{pmatrix}
         \dot{\tilde{u}}
        \\
         \dot{\tilde{v}}
    \end{pmatrix}
    &=
    i\,
    \mathbf{S}
    \mathbf{J}_\mathrm{f}
    \begin{pmatrix}
         \tilde{u}
        \\
         \tilde{v}
    \end{pmatrix}
    \nonumber\\
    \Rightarrow i
    \begin{pmatrix}
         \dot{\alpha}
        \\
         \dot{\alpha}^*
    \end{pmatrix}
    &=
    \underbrace{i
    \mathbf{S}
    \mathbf{J}_\mathrm{f}
    \mathbf{S}^{-1}}_{\equiv \mathbf{D}}
    \begin{pmatrix}
         \alpha
        \\
         \alpha^*
    \end{pmatrix}\, .
\end{align}
We multiplied both sides of Eq.~\eqref{eq:S_transform} with a complex factor $i$ in order to bring Eq.~\eqref{eq:DynamicMatrix} into a form where the so-called dynamic matrix $\mathbf{D}$ \cite{Xiao2009theory} of the system can be directly read out. This is crucial as the symplectic norm defined in Ref.~\cite{Soriente_2021} is formulated in terms of the eigenvectors of the dynamic matrix. 

The eigenvectors $\mathbf{v}_\pm$ of the dynamic matrix $\mathbf{D}$ can be calculated using the eigenvectors $\mathbf{w}_\pm$ of the Jacobian $\mathbf{J}_\mathrm{f}$:
\begin{align}
    \mathbf{J}_\mathrm{f} \mathbf{w}_\pm&=\mu_\pm \mathbf{w}_\pm
    \nonumber\\
    \Rightarrow i\, \mathbf{S}\mathbf{J}_\mathrm{f}\mathbf{S}^{-1}\mathbf{S}\mathbf{w}_\pm&=i\, \mu_\pm \mathbf{S} \mathbf{w}_\pm
    \nonumber\\
    \Rightarrow i\, \mathbf{D}\mathbf{S}\mathbf{w}_\pm&=i\, \mu_\pm \mathbf{S} \mathbf{w}_\pm
    \nonumber\\
    \Rightarrow \mathbf{D}\left(\sqrt{\frac{m\omega_0}{\hbar}}\mathbf{v}_\pm\right)&=i\mu_\pm\left(\sqrt{\frac{m\omega_0}{\hbar}}\mathbf{v}_\pm\right)\,.
\end{align}
In the last line, we used the definition of $\mathbf{D}$ from Eq.~\eqref{eq:DynamicMatrix} and introduced an additional factor of $\sqrt{m\omega_0/\hbar}$ to ensure that $|\mathbf{w}_\pm|=|\mathbf{v}_\pm|$, accounting for the fact that $\mathbf{S}$ is not a unitary transformation, cf. Eq.~\eqref{eq:SNotUnitary}. 

Since $\mathbf{w}_+ = \mathbf{w}_-^*$, this implies that $\mathbf{v}_+ = \mathbf{v}_-^*$, and we have $\mathbf{w}_+^\dagger=\mathbf{w}_-^T$, as well as $\mathbf{v}_+^\dagger=\mathbf{v}_-^T$ (with $T$ denoting transpose), allowing us to rewrite the classical symplectic norm defined in Eq.~\ref{eq:SymplecticNormDefinition} as
\begin{align}
\label{eq:SymplecticNormSorienteFormv+}
    ds^2 &= \left(w_{-,1}, w_{-,2}\right)
    \begin{pmatrix}
        0 & -i\\
        i & 0
    \end{pmatrix}
    \begin{pmatrix}
        w_{+,1}
        \\
         w_{+,2}
    \end{pmatrix}
    \nonumber\\
    &= \left(w_{-,1}, w_{-,2}\right)
    \mathbf{S}^\dag
    \left(\mathbf{S}^\dag\right)^{-1}
    \begin{pmatrix}
        0 & -i\\
        i & 0
    \end{pmatrix}
    \mathbf{S}^{-1}
    \mathbf{S}
    \begin{pmatrix}
        w_{+,1}
        \\
         w_{+,2}
    \end{pmatrix}
    \nonumber\\
    &= \left(
    \mathbf{S}
    \mathbf{w}_+\right)^\dag
    \left(\mathbf{S}^\dag\right)^{-1}
    \begin{pmatrix}
        0 & -i\\
        i & 0
    \end{pmatrix}
    \mathbf{S}^{-1}
    \mathbf{S}
    \mathbf{w}_+
    \nonumber\\
   &=
    \mathbf{v}_+^\dag
    \mathbf{I}_-
    \mathbf{v}_+\, ,
\end{align}
with $\mathbf{I}_-=\mathrm{diag}(1,-1)$, and where we used $\mathbf{v}_+ = \mathbf{S} \mathbf{w}_+$. Since the symplectic norm is real valued, one could also calculate it using $\mathbf{v}_-$, yielding the same result:
\begin{align}
\label{eq:SymplecticNormSorienteFormv-}
    ds^2=\left(ds^2\right)^\dag=\left( \mathbf{v}_+^\dag
    \mathbf{I}_-
    \mathbf{v}_+\right)^\dag
    =\mathbf{v}_-^\dag
    \mathbf{I}_-
    \mathbf{v}_-\,.
\end{align}
Equations~\eqref{eq:SymplecticNormSorienteFormv+} and \eqref{eq:SymplecticNormSorienteFormv-} show that the (classical) symplectic norm defined in Eq.~\eqref{eq:SymplecticNormDefinition} coincides exactly with the symplectic norm introduced in Ref.~\cite{Soriente_2021}. As the sign of Eq.~\eqref{eq:LinearizedModelFrequency} is determined by $ds^2$, the rotational sense of a ringdown is directly linked with the sign of the symplectic norm of the attractor. Since the sign of the symplectic norm of an attractor is positive (negative) for a minimum (maximum) of the underlying potential \cite{Soriente_2020, Soriente_2021}, the rotational sense of the ringdown can be used to differ between maxima and minima in the rotating frame potential.

\subsection{Spectral Response}

Having established a link between the rotational sense of ringdowns and the symplectic norm, we derive the `spectral response' of fluctuations around an attractor. This spectral response also allows us to extract the symplectic norm from ringdown measurements.

In analogy to driven-dissipative quantum systems~\cite{Soriente_2020, Soriente_2021}, we first introduce the correlator function
\begin{align}
    G^c(t'-t)&=\Theta(t'-t)\langle\tilde v(t) \tilde u(t')-\tilde u(t)\tilde v(t')\rangle_t \label{app:eq-green} \\ 
    &=\Theta(t'-t)\frac{2ds^2 |\zeta|^2}{\Gamma} e^{-\frac{\Gamma}{2}(t'-t)} \sin(\omega_\mathrm{lin} (t'-t)) \nonumber  ,
\end{align}
where `cc.' denotes the complex conjugate, and $\Theta(t-t')$ is the Heavyside step-function. The function $G^c$ correlates the displaced quadratures $\tilde{u}(t)$ and $\tilde{v}(t')$ at different times to capture the rotational sense of a ringdown near an attractor. This correlator is  reminiscent of Eq.~\eqref{eq:SymplecticNormDefinition}, and is analogous to a retarded Green's function in driven-dissipative quantum systems,  correlating at different times the creation and annihilation of bosons on top of a stable solution. Note that the second equality in Eq.~\eqref{app:eq-green} holds in the limit of the linearization procedure presented earlier.

The sign of the symplectic norm can now be read out directly by inspecting the peaks and dips of the spectral function~\cite{Soriente_2020, Soriente_2021, scarlatella2019spectralfunction}. This spectral response is related to the imaginary part of the Fourier transform of the Green's function $G^c(\omega)$ via
\begin{align}\label{app:eq-A}
    \mathcal A(\omega) &= -2\mathrm{Im}[G^c(\omega)].
\end{align}
Explicitly computing Eq.~\eqref{app:eq-A} using Eq.~\eqref{app:eq-green} yields the spectral response of our system:
\begin{align}
    \mathcal A(\omega)      = \frac{|\zeta|^2 ds^2}{2} \left[\frac{1}{(\omega-\omega_\mathrm{lin})^2+\frac{\Gamma^2}{4}}-\frac{1}{(\omega+\omega_\mathrm{lin})^2+\frac{\Gamma^2}{4}}\right] \, . 
\end{align}
A peak (dip) of $\mathcal{A}$ at positive frequencies therefore corresponds to a positive (negative) symplectic norm and hence to a minimum (maximum) of the rotating frame potential $H_\mathrm{rot}$ at $(u_\mathrm{f},v_\mathrm{f})$.

\section{Details of the Hamiltonian Reconstruction}\label{App:Hamiltonian_Reconstruction}

In this section, we provide further details on the Hamiltonian reconstruction process. We start the reconstruction by using a single ringdown measurement with the quadratures $u_j(t_j)$ and $v_j(t_j)$ measured at discrete times $t_j$. Assuming that our device is linearly damped, and using the extracted damping $\Gamma$, we calculate the change in the Hamiltonian $\Delta H_\mathrm{rot}(u_j, v_j)$ at each coordinates along the ringdown path by numerically integrating Eq.~\eqref{eq:Delta-H_rot}.

We then find the attractor $k$, e.g. $k = {1,2,3}$ for the case of three attractors, in which the resonator rings down into by looking at the final value $(u_\mathrm{f}, v_\mathrm{f})_k$. We fix the Hamiltonian $H_\mathrm{rot}(u_{\mathrm{f}}, v_{\mathrm{f}}) = C_k$ at the final coordinate $(u_{\mathrm{f}}, v_\mathrm{f})_k$ to an arbitrary value $C_k$. To do so, we add a constant offset to the Hamiltonian change $\Delta H_\mathrm{rot}$ calculated for the ringdown. This will insure that the Hamiltonian is singled-valued at each attractor, i.e., $H_\mathrm{rot}(u_{\mathrm{f}}, v_{\mathrm{f}})$ is the same for each ringdown ending up in the same attractor. At this stage, we simply set $C_k = 0$.

\begin{figure}[t]
	\centering
	\includegraphics[width=0.85\columnwidth]{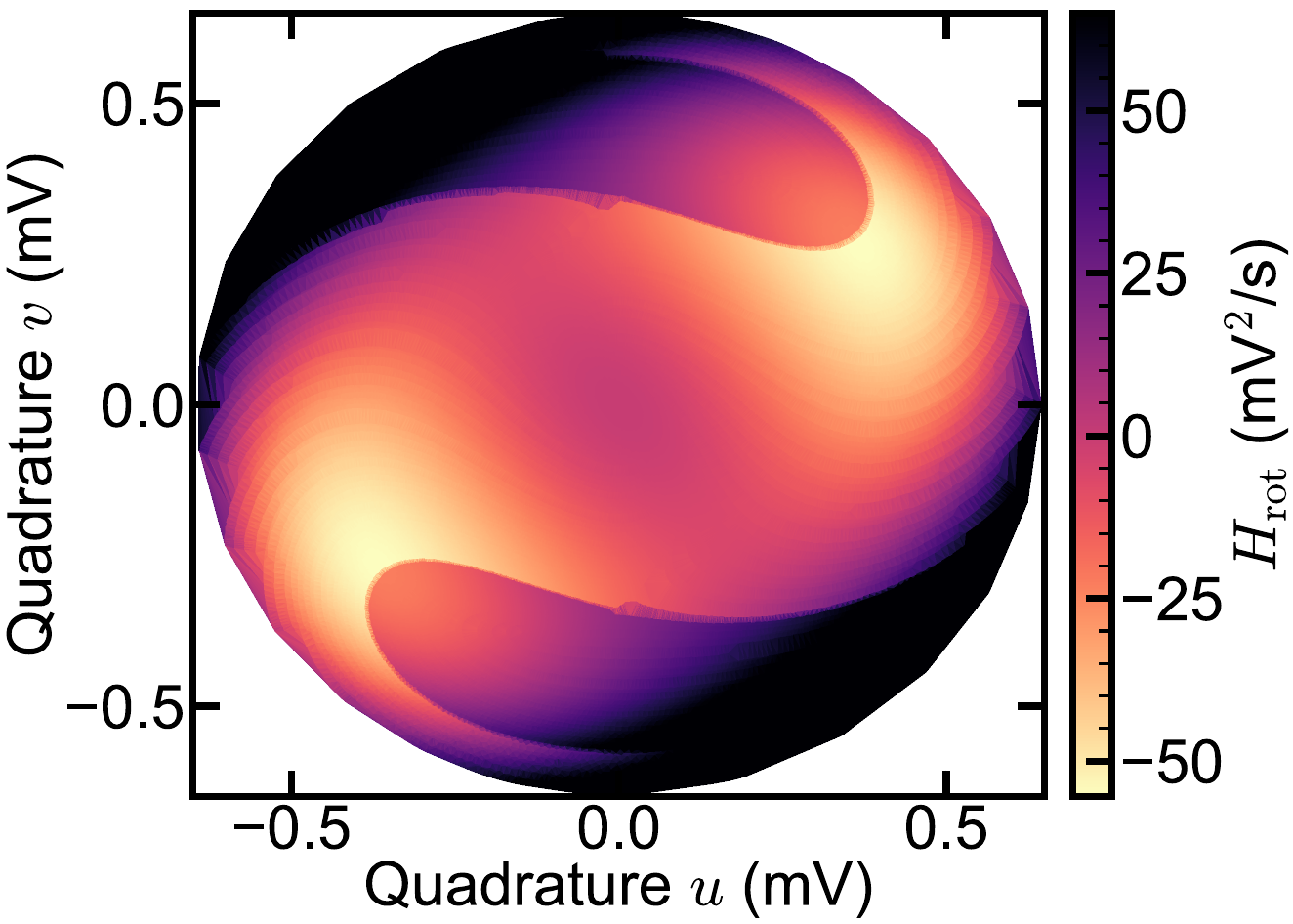}
	\caption{Hamiltonian reconstruction before adjusting the offset $C_k$ of each attractor. After using Eq.~\eqref{eq:Delta-H_rot} to recover the relative Hamiltonian potential along each ringdown trajectory and setting the end point of each ringdown to $C_k =0$, we obtain a discontinuous Hamiltonian. The clear discontinuity happens at the separatrix of the system since the offset between the different attractors $C_k-C_{k+1}$ has yet to be found.}
	\label{Fig_app:3} 
\end{figure}

As a second step, we repeat the integration procedure for all ringdown  measurements and check if the final value of each ringdown $(u_{\mathrm{f}}, v_\mathrm{f})_k$ is the same as the first ringdown, or if it ends up at another attractor. If it is another attractor, we label the new attractor as $k+1$, and add an arbitrary value $C_{k+1}$ to the trace, which we will later adjust. We repeat this procedure for all measured ringdowns, allowing to find the Hamiltonian change for all the ringdown paths, as shown in Fig.~\ref{Fig_app:3} for an example with three attractors.

 We generally find discontinuities in the Hamiltonian reconstruction at the separatrices between different attractors, see Fig.~\ref{Fig_app:3}. Such discontinuities are unphysical and stem from our choice to set all offsets to $C_k =0$. As a last step of the reconstruction, we thus need to find the relative height $C_k-C_{k+1}$ between the attractors such that the discontinuities disappear. In practice, we fix one of the constants (e.g., $C_1 = 0$), and vary the constant offsets $C_k$ of the other attractors ($C_2$ and $C_3$ in Fig.~\ref{Fig_app:3}) until the Hamiltonian is smooth (i.e., such that the second differentials of $H_\mathrm{rot}$ with respect to $u$ and $v$ are minimized). We furthermore set the minimum of $H_\mathrm{rot} = 0$, yielding Fig.~\ref{Fig:3}(a)(iii). For our demonstration examples, this optimization was done manually.

%%%%%%%%%%%%%%%%%%%%%%% References %%%%%%%%%%%%%%%%%%%%%%%%%
\addcontentsline{toc}{section}{References}
\bibliographystyle{bibstyle-jack}
\bibliography{errthing.bib}

\end{document}